\documentclass[showpacs,prb,twocolumn]{revtex4}
\usepackage{graphicx}
\usepackage{dcolumn}
\usepackage{bm}
\usepackage{amsfonts}
\usepackage{amsmath}
\usepackage{amssymb}

\begin{document}

\title{Bipolar charge-carrier injection in
semiconductor/insulator/conductor heterostructures: self-consistent consideration}
\author{S.V.~Yampolskii}
\email{yampolsk@tgm.tu-darmstadt.de}
\author{Yu.A.~Genenko}
\author{C.~Melzer}
\author{K.~Stegmaier}
\author{H.~von Seggern}
\affiliation{Institute of Materials Science, Darmstadt University of Technology,
Petersenstra{\ss}e 23, D-64287 Darmstadt, Germany}
\date{\today}
\begin{abstract}
A self-consistent model of bipolar charge-carrier injection and
transport processes in a semiconductor/insulator/conductor system is developed
which incorporates space-charge effects in the description of the
injection process. The amount of charge-carriers injected 
is strongly determined by the energy barrier emerging at the
contact, but at the same time the electrostatic potential generated
by the injected charge-carriers modifies the height of this
injection barrier itself. In our model, self-consistency is obtained
by assuming continuity of the electric displacement and of the
electrochemical potential all over the system. The constituents of
the system are properly taken into account by
means of their respective density of state distributions. The
consequences resulting from our model are discussed on the basis of
an indium tin oxide/organic semiconductor/conductor
structure. The distributions of the charge carriers and the electric
field through the electrodes and the organic layer are calculated. The
recombination- and current-voltage characteristics are analyzed for different heights
of injection barriers and varying values of the recombination rate and compared
with the measured current-voltage dependences for an indium tin oxide/poly(phenylene vinylene)/Ca
structure. The voltage dependences of the recombination efficiency for
the different values of injection barriers
and recombination rate reveal optimum conditions for the device performance.

\end{abstract}
\maketitle

\section{Introduction}
Though the problem of bipolar charge injection in insulator media is
many years old~\cite{inject1,inject2,inject3,inject4}, it anew
became of great interest during last years in view of application of
insulating materials as basic elements of electronic devices such as
ferroelectric random access memories (FeRAMs)~\cite{Dawber}, organic
field-effect transistors (OFETs) or organic light-emitting diodes
(OLEDs)~\cite{Blom2000,Walker2002,Spintronics,JAPReview2007}. In the
latter devices organic semiconductors are used which show many
properties of dielectric materials, especially relatively large band
gaps and thus, a low intrinsic charge carrier density. The charge
carriers have to be injected into the organic layer from the
electrodes and thereby must overcome the injection barriers at the
organic/electrode interfaces. It is experimentally established that the
injection conditions influence significantly the OLED
performance~\cite{exp1,exp2,exp3,exp4}.

Different models for the injection process are proposed in the
literature. For low injection barriers, one expects the contact to
be Ohmic, meaning that the contact is able to supply more charges
per unit time than the bulk of the insulator can transport. In this
case, a space-charge region is formed and the electric field at the
interface is supposed to vanish~\cite{Lampert}. Because excess
charge-carriers dominate the charge transport in insulators, a
space-charge-limited current (SCLC) density of the form $j\sim
V^2/L^3$ is observed (in the absence of charge-carrier traps), where
$L$ is the sample thickness and $V$ is the applied voltage.
Current-voltage ({\it IV}) characteristics of space-charge limited
devices are determined by the bulk properties of the insulator with
no influence of the contact properties
\cite{Blom1996APL,Blom1996PRB,SyntMet05}.

For high injection barriers, one anticipates the injection rate
across the conductor/insulator interface to dominate the {\it IV}
characteristic of the system. The models to describe injection are
the Fowler-Nordheim (FN) tunneling model or the Richardson-Schottky
(RS) model for thermionic injection~\cite{Sze}. The FN model assumes
tunneling through a triangular barrier into an unbound continuum of
states. The RS model on the other hand describes charge injection as
a thermally activated hopping over the potential barrier, where
barrier lowering due to the superposition of the external
electrostatic potential and the image-charge potential is
considered. An attempt to incorporate space-charge effects into the hopping injection 
model was undertaken in Refs.~\onlinecite{hopping1,hopping2}. 
However, all these models consider injection as a
single electron process.

An alternative description of the injection process is given by the
drift-diffusion theory involving electron-electron interaction in a
mean-field approximation~\cite{Walker2002,Martin2005}. In this
homogeneous continuum model widely used for the description of
conventional crystalline semiconductors, the drift-diffusion
equation in combination with the Poisson equation involves
space-charge effects, but meets the problem of self-consistency in
the boundary conditions. Namely, the electrostatic potential
generated by the injected charge carriers modifies the injection
barrier, but on the other hand the amount of charge-carriers
injected per unit time depends on the barrier height. Thus, the
values of the electrical field and charge-carrier density at the
interface cannot be imposed but have to be found self-consistently
implying the modification of the injection barrier, too. 
An effort to perform a self-consistent analysis of the contact phenomena
was made in Ref.~\onlinecite{Tessler1}. However, the current and the field 
in this approach were not defined in a self-consistent way.
This can be
performed if the drift-diffusion theory involves both, the insulator
and the conductor side of the system, in analogy to a strongly
asymmetric {\it pn}-junction. In the work of Neumann {\it et
al}.~\cite{JAP100}, a self-consistent numerical treatment of the
unipolar injection and transport processes in a
conductor/insulator/conductor device was presented in which
continuity of the electrochemical potential and the electric
displacement was assumed everywhere in the system, in particular at
the contacts. The conductor and the insulator were characterized by
their specific density of state (DOS) distributions. The exact
analytical solution in a self-consistent manner for the charge
injection across the sole conductor/insulator interface has been
also obtained~\cite{PRB2007}.

This model does not yet allow for a self-consistent description of
organic light emitting diodes, where bipolar transport takes
place~\cite{Walker2002,JAPReview2007,JCScott1,JCScott2}. To do so,
one has to account for the recombination of charge carriers (holes
and electrons) which are usually present in the organic constituent
of the heterostructure under consideration. In such systems the
recombination is often believed to be of Langevin
type~\cite{recomb1, recomb2}, {\it i.e.} is considered as a
bimolecular process. Many OLED models involve the recombination of
electrons and holes (see, for example,
Refs.~\onlinecite{Blom2000,SyntMet05,Walker2002,Shen1998,BlomAPL97,BlomIEEE,Blom,Walker2000,Martin2}).
Probably, the most complete description was presented by Malliaras
and Scott~\cite{Malliaras1,Malliaras2}, who included a surface 
recombination at the contacts as well as a bimolecular recombination
kinetics and diffusion in the bulk. However, the assumed boundary conditions
for Ohmic contacts excluded diffusion at the contacts, where it is
most important. A comprehensive one-dimensional numerical model
accounting for space-charge effect was developed by Tuti\v{s} {\it
et al.}~\cite{Tutis2001} which comprised the hopping transport in
single and bilayer devices and tunnel injection from electrodes.
However, besides the injection barriers given by the bare difference
of the chemical potential in the metal and the frontier molecular
orbital in the organic material this model involves a range of
artifacts such as tunneling factors, effective attempt frequencies,
etc. Hence, in spite of a good agreement with experiments, this
numerical tool is very complicated.

In the present paper, we apply the diode model given in Ref.~\onlinecite{JAP100}
to elaborate a self-consistent OLED model. The
properties of a semiconductor/organic/conductor structure are
modeled where holes and electrons are simultaneously injected from an
indium tin oxide (ITO) anode and a Ca cathode into the organic layer, respectively. The
focus will be on the influence of the injection barriers
and the recombination on the spatial distribution of the injected
charge carriers and on the resulting {\it IV}
characteristics as well as recombination efficiencies. Therefore, for simplicity, trap states and a field
dependent mobility of the organic semiconductor constituent will be
excluded at this stage.

The paper is organized as follows. In Sec.~II we present the
theoretical model of the charge injection in
semiconductor/insulator/conductor structure. Then, using specific
material parameters, we model the spatial distributions of the
charge-carrier density and of the electric field (Sec.~IIIA). In Sec.~IIIB, the
{\it IV} characteristics of the system (Sec.~IIIB) are
calculated varying the heights of injection barriers and the rate of
the Langevin recombination. We also discuss the calculated voltage
dependences of the recombination efficiency of the considered
structure (Sec.~IIIC). Our model is applied to a measured {\it IV}
characteristic of an OLED composed of a poly(phenylene
vinylene) (PPV) organic single layer sandwiched between ITO and Ca
electrodes (Sec.~IIID). Finally, in Sec.~IV our results are
summarized.

\section{Model}

\begin{figure*}[!htbp]
\includegraphics[width=12cm]{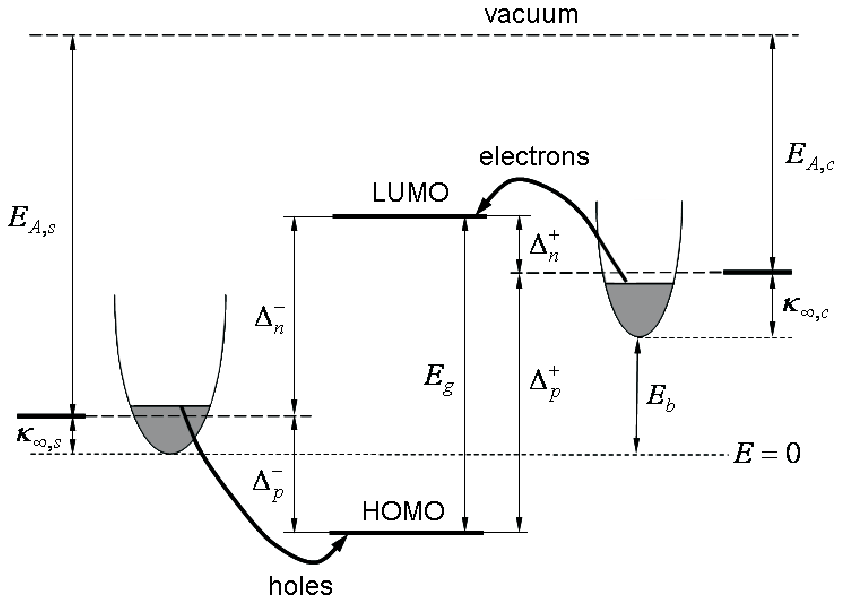}
\caption{Schematic band diagram of the considered semiconductor/insulator/conductor structure.} \label{fig0}
\end{figure*}

Let us consider an insulator of thickness $L$ sandwiched in between
a semiconductor and a conductor electrode. The insulator is supposed
to be extended over the space with $-L/2\leq x\leq L/2$, whereas the
semiconductor and conductor electrodes are extended over the
half-spaces with $x<-L/2$ and $x>L/2$, respectively. We assume that
the semiconductor electrode injects holes into the insulator layer
while the conductor electrode injects electrons. The band structure of the system under consideration is shown
schematically in Fig.~\ref{fig0}.

\subsection{Electrodes}

Here we consider ITO as the hole-injecting electrode being an electron-conducting
semiconductor with deep laying conduction band~\cite{Martin2005,JAPReviewITO}. 
In the Thomas-Fermi approximation, one can
deduce the electrochemical potential $\kappa _{s}$ of the
hole-injecting electrode as a function of the spatial coordinate
$x$,
\begin{align}
\kappa _{s}\left( x\right) = \frac{\hbar ^{2}}{2m_s^*}\left[ 3\pi
^{2}n_s\left( x\right) \right] ^{2/3}-e\phi \left( x\right), \label{kappa1}
\end{align}
\noindent where $n_s\left( x\right) $ is the electron density in the electrode, $m_s^*$ 
is the effective mass of the charge carriers in the semiconductor, $\hbar$ is Planck's constant,
$e$ is the elementary charge, and $\phi \left( x\right)$ 
is the electrostatic potential (notice that the energy level $E=0$
coincides with the bottom of the semiconductor conduction band).

In general, everywhere in the heterostructure the electrochemical potential 
$\kappa \left( x\right) $ relates the steady-state current density $j$ with
the charge-carrier density. For a one-dimensional geometry, the current
remains constant across the whole space and $j$ is given by the conductivity
$\sigma $ and the derivative of $\kappa \left( x\right) $.

Accordingly, for the hole-injecting electrode,
\begin{align}
j=\mu_{s} n_s \left( x \right) \frac{d\kappa _{s}\left( x\right) }{dx}, \label{current1}
\end{align}
\noindent where $\mu _{s}$ is the charge-carrier mobility in the semiconducting electrode. The electric field
$F_{s}\left( x\right) $ in the electrode obeys Gauss law,
\begin{align}
F_{s}^{\prime }\left( x\right) = -\frac{e}{\epsilon _{s}\epsilon _{0}}\delta n_s\left( x\right),  \label{Gauss1}
\end{align}
\noindent where $\epsilon _{s}$ is the relative permittivity of the
electrode, $\delta n_s\left( x\right) =n_s\left(x\right) -n_{\infty,s }$ is the excess electron density near the
interface with respect to the electron density in the
conduction band at infinite distance from the semiconductor/insulator
interface, $n_{\infty,s }$. The value of this excess density is supposed to be small in comparison with the
background electron density, $\left\vert \delta n_s\left( x\right) \right\vert \ll n_{s,\infty }$. 
Hence, the linearized Thomas-Fermi approximation can be
applied, leading to a differential equation for $F_{s}$,
\begin{align}
\frac{j}{\sigma _{s}}=-l_{TF,s}^{2}F_{s}^{\prime \prime }\left(x\right) +F_{s}\left( x\right), \label{Feq1}
\end{align}
\noindent with the conductivity in the electrode $\sigma _{s}=e\mu _{s}n_{\infty,s }$ and
\begin{align}
l_{TF,s}=\sqrt{\frac{2}{3}\frac{\epsilon _{s}\epsilon _{0} \kappa _{\infty,s} }{e^{2}n_{\infty,s }}},  \label{TF1}
\end{align}
\noindent being the Thomas-Fermi screening length of the
electrode. Here, $\kappa _{\infty ,s}$ is the Fermi level of electrons
with respect to the bottom of the conduction band.

Since the space-charge zone in the electrode is of finite thickness,
the gradient of $F_{s}\left( x\right) $ has to vanish at 
infinite distance from the contact, and the solution for the
electric field in the electrode reads
\begin{align}
F_{s}\left( x\right) =\frac{j}{\sigma _{s}}+\left[ F_{s}\left( -\frac{L}{2}\right) 
-\frac{j}{\sigma _{s}}\right] \exp \left( \frac{x+L/2}{l_{TF,s}}\right),   \label{field1}
\end{align}
\noindent where the field in the electrode at the
semiconductor/insulator interface, $F_{s}\left(-L/2\right) $, is unknown and has to be determined by the
boundary conditions.

The distribution of the electric field in the electron-injecting
 electrode (conductor) is found in the same way as above.
Even though this derivation was performed for a 
semiconductor it may be still meaningful for simple metals~\cite{Kohn}.
The electrochemical potential in the conductor reads
\begin{align}
\kappa _{c}\left( x\right) = \frac{\hbar ^{2}}{2m_c^*}\left[ 3\pi^{2}n_c\left( x\right) \right]^{2/3} 
-e\phi \left( x\right) +E_{b},  \label{kappa3}
\end{align}
\noindent where $n_c\left( x\right) $ is the electron density in the
electrode, $m_c^*$ is the effective mass in the conductor and $E_{b}$ is
the bottom of the conduction band. 
Performing calculations in the similar manner as for the hole-injecting electrode, we 
obtain the electric field in the conductor electrode, 
\begin{align}
F_{c}\left( x\right) =\frac{j}{\sigma _{c}}+\left[ F_{c}\left( \frac{L}{2}\right) 
-\frac{j}{\sigma _{c}}\right] \exp \left( -\frac{x-L/2}{l_{TF,c}}\right),   \label{field3}
\end{align}
\noindent with the unknown electric field $F_{c}\left( L/2\right) $ at the insulator/conductor 
interface again being determined by the boundary conditions, the conductivity 
$\sigma _{c}=e\mu _{c}n_{\infty,c }$ and the Thomas-Fermi screening length
\begin{align}
l_{TF,c}=\sqrt{\frac{2}{3}\frac{\epsilon _{c} \epsilon _{0} \kappa _{\infty ,c} }{%
e^{2}n_{\infty,c }}}.  \label{TF3}
\end{align}
Here $\mu _{c}$ is the electron mobility of the metal electrode, $\kappa_{\infty ,c}$ 
is the Fermi level of electrons with respect to
the conduction band bottom, $n_{\infty,c }$ is the background electron density 
in the conduction band at infinite distance from the conductor/insulator
interface and $\epsilon_c$ is the relative permittivity of the conductor.

\subsection{Insulator}

The energetic differences between the Fermi levels $\kappa_{\infty}$ in the
electrodes and the bottom of
the conduction band or the top of the valence band in the insulator are
defined as the injection barriers $\Delta _{n}$ and $\Delta _{p}$ for
electrons and holes, respectively. These barriers relate to the difference
between the electrode work functions $E_{A}$ as well as to the insulator
interband gap energy $E_{g}$:
\begin{align}
E_{A,s}-\Delta _{n}^{-}=E_{A,c}-\Delta _{n}^{+}, \label{barrier1}
\end{align}
\begin{align}
\Delta_{n}^{-}+\Delta _{p}^{-}=\Delta _{n}^{+}+\Delta _{p}^{+}=E_{g}.  \label{barrier2}
\end{align}
From now on, the minus and plus superscripts denote the barriers at the interfaces $x=-L/2$ and $x=L/2$,
respectively.

Let us characterize the insulator by the DOS functions,
$g_{n,i}\left( E\right) $ and $g_{p,i}\left( E\right) $, describing
extended states in the conduction and valence bands in which
electron and hole transport takes place, respectively. Due to the
large value of the energy gap, it is possible to introduce two
separate electrochemical potentials; $\kappa _{n,i}$ for electrons and $%
\kappa _{p,i}$ for holes. Introducing band edges means that the DOS
function $g_{n,i}\left( E\right)=0 $ when $E<0 $ and the DOS
function $g_{p,i}\left( E\right)=0 $ when $E>0 $. Hence, the
densities of electrons and holes in the extended states can be
calculated using Boltzmann statistics, thus allowing solely for a
proper description of non-degenerate insulators,
\begin{align}
n_{i}\left( x\right) =\int\limits_{-\infty}^{\infty }g_{n,i}\left[ E-\Delta _{n}^{-}-\kappa _{\infty ,s}
+e
\phi\left( x \right)\right]
f_{n}\left( E\right) dE,  \label{nisol}
\end{align}
\begin{align}
p_i \left( x \right) = \int\limits^{\infty}_{-\infty} g_{p,i} \left[ E+ \Delta_p^- -
\kappa_{\infty,s}+e\phi\left( x \right) \right] f_p \left( E \right) dE,  \label{pisol}
\end{align}
\begin{align}
f_n \left( E \right)= \exp \left[ \frac{\kappa_{n,i} \left( x \right)-E}{kT} \right],  \label{Fermi1}
\end{align}
\begin{align}
f_p \left( E \right)= \exp %
\left[ \frac{E - \kappa_{p,i} \left( x \right)}{kT} %
\right],   \label{Fermi2}
\end{align}
here $T$ is the absolute temperature and $k$ is the Boltzmann constant.

The electrochemical potentials $\kappa _{n,i}$ and $\kappa _{p,i}$ are consequently
expressed in terms of the charge-carrier densities $n_{i}$\ and $p_{i}$,
\begin{align}
\kappa _{n,i}\left( x\right) =kT\ln \left[ \frac{n_{i}\left( x\right) }{%
N}\right] +\Delta _{n}^{-}+\kappa _{\infty ,s}-e\phi \left(
x\right),  \label{kappa2n}
\end{align}
\begin{align}
\kappa _{p,i}\left( x\right) =-kT\ln \left[ \frac{p_{i}\left( x\right) }{%
P}\right] -\Delta _{p}^{-}+\kappa _{\infty ,s}-e\phi \left(
x\right),    \label{kappa2p}
\end{align}
where the quantities
\begin{align}
N=\int\limits_{0}^{\infty }g_{n}\left( E^{\prime }\right) \exp
\left( -\frac{E^{\prime }}{kT}\right) dE^{\prime },  \label{Ni}
\end{align}
\noindent and
\begin{align}
P=\int\limits_{-\infty }^{0}g_{p}\left( E^{\prime }\right) \exp
\left( \frac{E^{\prime }}{kT}\right) dE^{\prime }   \label{Pi}
\end{align}
can be understood as the effective total densities of states
available in the conduction and valence bands of the insulator,
respectively.

The total current density in the insulator consists of the electron and hole contributions
and reads:
\begin{align}
j = j_p \left( x \right) + j_n \left( x \right), \label{curtot}
\end{align}
\noindent with
\begin{align}
j_n = \sigma _{n,i}\left( x\right) F_{i}\left( x\right) +
kT \mu_{n,i}n_{i}^{\prime }\left( x\right), \label{currentn2fin} \\
j_p= \sigma_{p,i}\left( x\right) F_{i}\left( x\right) -kT \mu _{p,i}p_{i}^{\prime }\left(
x\right),  \label{currentp2fin}
\end{align}
\noindent where $\sigma _{n,i} = e \mu_{n,i} n \left( x \right)$ and
$\sigma _{p,i} = e \mu_{p,i} p \left( x \right)$ are the
conductivities of electrons and holes, respectively, with the
respective electron and hole mobilities $\mu_{n,i}$ and $\mu_{p,i}$.
When accounting for the electron-hole recombination in the insulator, the steady-state
continuity equations for holes and electrons read
\begin{align}
\frac{dj_n \left( x \right)}{dx} = eBR \left[ n_i \left( x \right) p_i \left(
x \right) - n_{\text{int}}^2 \right], \label{difkappa1}
\end{align}
\begin{align}
\frac{dj_p \left( x \right)}{dx} = -eBR
\left[ n_i \left( x \right) p_i \left(
x \right) - n_{\text{int}}^2 \right],  \label{difkappa2}
\end{align}
\noindent where $0 \le R \le 1$ is the recombination parameter, $B$ is the
Langevin recombination coefficient,
\begin{align}
B = \frac{e}{\epsilon_i \epsilon_0} \left( \mu_{n,i} + \mu_{p,i} \right), \label{BLang}
\end{align}
\noindent and the intrinsic charge density $n_{\text{int}}$ is given
by $n_{\text{int}}^2=N P \exp \left( -E_g/kT
\right)$~\cite{Sze,Walker2000}. We note one more time that, in steady
state, the total current remains constant through the entire system.

Finally, from Eqs.~(\ref{currentn2fin})-(\ref{difkappa2}) we obtain the system of equations for the charge-carrier densities
and the electric field, governed by Gauss law, in the insulator:
\begin{align}
& n_{i}^{\prime }\left( x\right) F_{i}\left( x\right) +n_{i}\left(
x\right) F_{i}^{\prime }\left( x\right) +\frac{kT}{e}n_{i}^{\prime
\prime }\left( x\right) \notag \\
& =\frac{B}{\mu _{n,i}}R \left[ n_i \left( x \right) p_i \left(
x \right) - n_{\text{int}}^2 \right],  \label{eq1i}
\end{align}
\begin{align}
& p_{i}^{\prime }\left( x\right) F_{i}\left( x\right) +p_{i}\left(
x\right) F_{i}^{\prime }\left( x\right) -\frac{kT}{e}p_{i}^{\prime
\prime }\left( x\right) \notag \\
& =-\frac{B}{\mu _{p,i}}R \left[ n_i \left( x \right) p_i \left(
x \right) - n_{\text{int}}^2 \right],   \label{eq2i}
\end{align}
\begin{align}
F_{i}^{\prime }\left( x\right) =\frac{e}{\epsilon _{i}\epsilon _{0}
}\left[ p_{i}\left( x\right) -n_{i}\left( x\right) \right].  \label{eq3i}
\end{align}

\subsection{Self-consistency and boundary conditions at the contacts}

Assuming no dipole layer or surface charge at the interfaces one has to require continuity of the
electrical displacement and of the electrochemical potential~\cite{Landau},
\begin{align}
\epsilon F\left( x\right) & =\text{continuous},   \label{bcD} \\
\kappa \left( x\right) & =\text{continuous}.  \label{bckappa}
\end{align}
This continuity may be provided self-consistently matching the expressions of the electrical displacements
and electrochemical potentials.
In particular, one can write
\begin{align}
\epsilon _{s}F_{s}\left( -\frac{L}{2}\right) =\epsilon _{i}%
F_{i}\left( -\frac{L}{2}\right) ,\quad \epsilon _{i}F%
_{i}\left( \frac{L}{2}\right) =\epsilon _{c} F_{c}\left( \frac{L}{2}\right). \label{bcF}
\end{align}
Matching the electrochemical potentials we obtain four nonlinear boundary conditions:
\begin{align}
\ln \left[ \frac{n_i \left( -L/2 \right)}{N%
} \right] +\frac{\Delta_n^-}{kT} + \frac{e l_{TF,s}}{kT} \left[ \frac{%
\epsilon_i}{\epsilon_s} F_i \left( -\frac{L}{2} \right) - \frac{j}{%
\sigma_s} \right] =0,   \label{bc1}
\end{align}
\begin{align}
\ln \left[ \frac{p_i \left( -L/2 \right)}{P%
} \right] +\frac{\Delta_p^-}{kT} - \frac{e l_{TF,s}}{kT} \left[ \frac{%
\epsilon_i}{\epsilon_s} F_i \left( -\frac{L}{2} \right) - \frac{j}{%
\sigma_s} \right] =0,   \label{bc2}
\end{align}
\begin{align}
\ln \left[ \frac{n_i \left( L/2 \right)}{N}
\right] +\frac{\Delta_n^+}{kT} - \frac{e l_{TF,c}}{kT} \left[ \frac{\epsilon_i}{\epsilon_c}
F_i \left( \frac{L}{2} \right) - \frac{j}{\sigma_c} \right] =0,
\label{bc3}
\end{align}
\begin{align}
\ln \left[ \frac{p_i \left( L/2 \right)}{P}
\right] +\frac{\Delta_p^+}{kT} + \frac{e l_{TF,c}}{kT} \left[ \frac{\epsilon_i}{\epsilon_c}
F_i \left( \frac{L}{2} \right) - \frac{j}{\sigma_c} \right] =0,
\label{bc4}
\end{align}
\noindent which depend on parameters of both the insulator and the
electrodes. Notice, that from Eqs.~(\ref{bc1}) and (\ref{bc4}) the
 following relations are obtained:
\begin{align}
n_{i}\left( -L/2\right) p_{i}\left( -L/2\right) & = n_{i}\left( L/2\right) p_{i}\left( L/2\right) \notag \\
& = N P\exp \left( -\frac{E_g}{kT} \right). \label{bc3dim}
\end{align}
As the last boundary condition, we can use the value of the current density
taken at one of the interfaces:
\begin{align}
j = j_p \left( -L/2 \right) + j_n \left( -L/2 \right).  \label{bc5dim}
\end{align}

\begin{table*}
\caption{\label{Materialparameters}Typical material parameters for electrodes and an organic semiconductor
(Refs.~\onlinecite{Martin2005,Ashcroft,Mergel2002,Mergel2004,Fujiwara2005}). 
The parameters are deduced assuming $T=300$~K and
$m_e$ is the free electron mass.}
\begin{ruledtabular}
\renewcommand{\arraystretch}{1.5}
\begin{tabular}{ccccccc|ccccc|cccccc}
\multicolumn{7}{c|}{ITO} & \multicolumn{5}{c|}{Organic semiconductor} &
\multicolumn{5}{c}{Ca} \vspace{2pt}\\
\hline
$l_{TF,s}$ & $n_{\infty,s} $ & $\epsilon_s $ & $ \mu_s $ & $m_{s}^*$ & $\kappa_{\infty,s} $ & $E_{A,s}$ &
${N,P} $ &  $\epsilon_i $ & $ \mu_{p,i} $ & $ \mu_{n,i} $ & $E_g$ &
$l_{TF,c}$ & $n_{\infty,c} $ & $ \mu_c $ & $ \kappa_{\infty,c} $ & $E_{A,c}$ \vspace{2pt}\\
($\mathring{\text A}$) & (cm$^{-3}$) &  & $\left( \displaystyle \frac{\text{cm}^2}{\text{V s}}\right)$ & ($m_e$) & (eV) & (eV) &
(cm$^{-3}$) &     &  $\left( \displaystyle \frac{\text{cm}^2}{\text{V s}}\right)$
& $\left( \displaystyle \frac{\text{cm}^2}{\text{V s}}\right)$ & (eV) &
($\mathring{\text A}$) & (cm$^{-3}$) &  $\left( \displaystyle \frac{\text{cm}^2}{\text{V s}}\right)$ & (eV) & (eV) \vspace{2pt}
\\
\hline
8.6 & 10$^{20}$ & 9.3 & 30 &  0.35 & 0.225 & 4.7 & 10$^{21}$ &  3  & 10$^{-4}$ & 10$^{-6}$ &  2.4 &
0.8 & $2.6 \cdot 10^{22}$ & 66.7 & 4.68 & 2.87
\end{tabular}
\end{ruledtabular}
\vspace{0.5cm}
\end{table*}

\section{Physical and numerical analysis}

The full set of nonlinear differential equations (\ref{eq1i})-(\ref{eq3i}) with the nonlinear boundary
conditions (\ref{bc1})-(\ref{bc4}) and (\ref{bc5dim}) has to be solved numerically.

As an example for an insulator, an organic semiconductor can be
considered, exhibiting many typical characteristics of insulators
such as relatively large band gaps up to 3~eV and, hence, the
absence of intrinsic charge carriers. In simple organic
light-emitting diodes (OLEDs) a thin layer of a organic
semiconductor is contacted with a low work-function metal and a high
work-function transparent conducting oxide. Here, we consider indium
tin oxide (ITO) as the hole-injecting electrode and Ca as the
electron-injecting contact. ITO is typically employed as anode in
OLEDs, since it provides a decent conductivity and a sufficient high
work function (up to 5~eV) to allow for efficient hole injection
while being transparent in the visible region of the light spectrum
to ensure light out-coupling. On the other hand, Ca with a
work function of about 2.9~eV delivers efficiently electrons and,
hence, is used as cathode in OLEDs. From now on, it is assumed that
the material specific quantities of the organic semiconductor and
the electrodes adopt the typical values given in
Table~\ref{Materialparameters}.

The injection barriers $\Delta_{p,n}$ are given by the energetic
difference between the highest occupied molecular orbital (HOMO) or
the lowest unoccupied molecular orbital (LUMO) in the organic
semiconductor and $\kappa_{\infty}$ of the respective electrodes.
Change of the barrier heights at the interfaces and of the gap 
energy in the organic semiconductor, leaving the electrodes unchanged, can be 
understood as considering different organic semiconductors. Here, for simplicity and 
without loss of generality, we assume that the gap energy in organic constituent 
is fixed. Therefore, changing the value of the barrier height at one interface 
results in equally shifted energies of the HOMO and LUMO levels which entails the change 
of the barrier at another interface.

Notice, that specifying the material parameters for the insulator and the electrodes leads to
substantial consequences for the boundary conditions.  
The current density $j$ is multiplied  by the small factors $l_{TF,s}/ kT\mu_s n_{\infty, s}$ 
in Eqs.~(\ref{bc1})-(\ref{bc2}) and
by the small factor $l_{TF,c}/ kT \mu_c n_{\infty, c}$ in Eqs.~(\ref{bc3})-(\ref{bc4}).
Hence, the boundary conditions do not depend directly on $j$ in most practical cases.
Under this approximation, equations~(\ref{bc2})-(\ref{bc3}) can be reduced to
\begin{align}
p_i \left( -\frac{L}{2} \right) & = {P} \exp \left[- \frac{\Delta_p^-}{kT} +
\frac{e l_{TF,s}}{kT} \frac{\epsilon_i}{\epsilon_s} F_i \left( -\frac{L}{2} \right)\right],
\label{bd1dimnul}  \\
n_i \left( \frac{L}{2} \right) & = {N} \exp \left[- \frac{\Delta_n^+}{kT} +
\frac{e l_{TF,c}}{kT} \frac{\epsilon_i}{\epsilon_c} F_i \left( \frac{L}{2} \right)\right].
\label{bd2dimnul}
\end{align}
As a consequence, the dependence of $F_i(\mp L/2)$ as
well as $p_i(- L /2)$ or $n_i(L /2)$ on the
current is only due to Eq.~(\ref{bc5dim}). Apparently, the
injection barriers effectively vary with $\propto F_i(\mp L /2)$
similarly to the case of unipolar carrier
injection~\cite{JAP100,PRB2007}. This change in injection barriers
leads to the definition of effective injection barriers,
\begin{align}
\label{deltapeff}
\Delta_{\mbox{eff}}^{-}&=\Delta_p^- - e \frac{\epsilon_i}{\epsilon_s} F_i \left( -\frac{L}{2} \right) l_{TF,s}\\
&=\Delta_p^- - eF_s \left( -\frac{L}{2} \right)l_{TF,s},  \label{deltapeffdim} \\
\label{deltaeeff}
\Delta_{\mbox{eff}}^{+}&=\Delta_n^+ - e \frac{\epsilon_i}{\epsilon_c} F_i \left( \frac{L}{2} \right) l_{TF,c}\\
&=\Delta_n^+ - eF_c \left( \frac{L}{2} \right)l_{TF,c}.
\label{deltaeeffdim}
\end{align}
The changes in the effective injection barriers correspond to the
amount of energy a charge carrier gains (or loses) in the electric
field prevailing in the electrodes. This change, as is seen from
Eqs.~(\ref{deltapeffdim}) and (\ref{deltaeeffdim}), is not
necessarily negative but can be positive, since a space-charge
region at the interface might serve as a potential barrier itself.
The electric field, however, cannot be arbitrary large. Assuming Boltzmann statistics
for both injected electrons and holes requires that the electrochemical 
potentials~(\ref{kappa2n}) and~(\ref{kappa2p}) do not approach the respective band edges so that 
inequalities $n_i \ll N $ ans $p_i \ll P $ hold.
Considering boundary conditions~(\ref{bd1dimnul})-(\ref{bd2dimnul}), this imposes a requirement
on the electric field at the interfaces so that both effective barriers~(\ref{deltapeff}) and~(\ref{deltaeeff})
remain positive. To keep the validity of Boltzmann statistics the above requirements are controlled during
the further numerical calculations.  

\subsection{Spatial distributions of charge carriers and electric field}

\begin{figure*}[!htbp]
\includegraphics{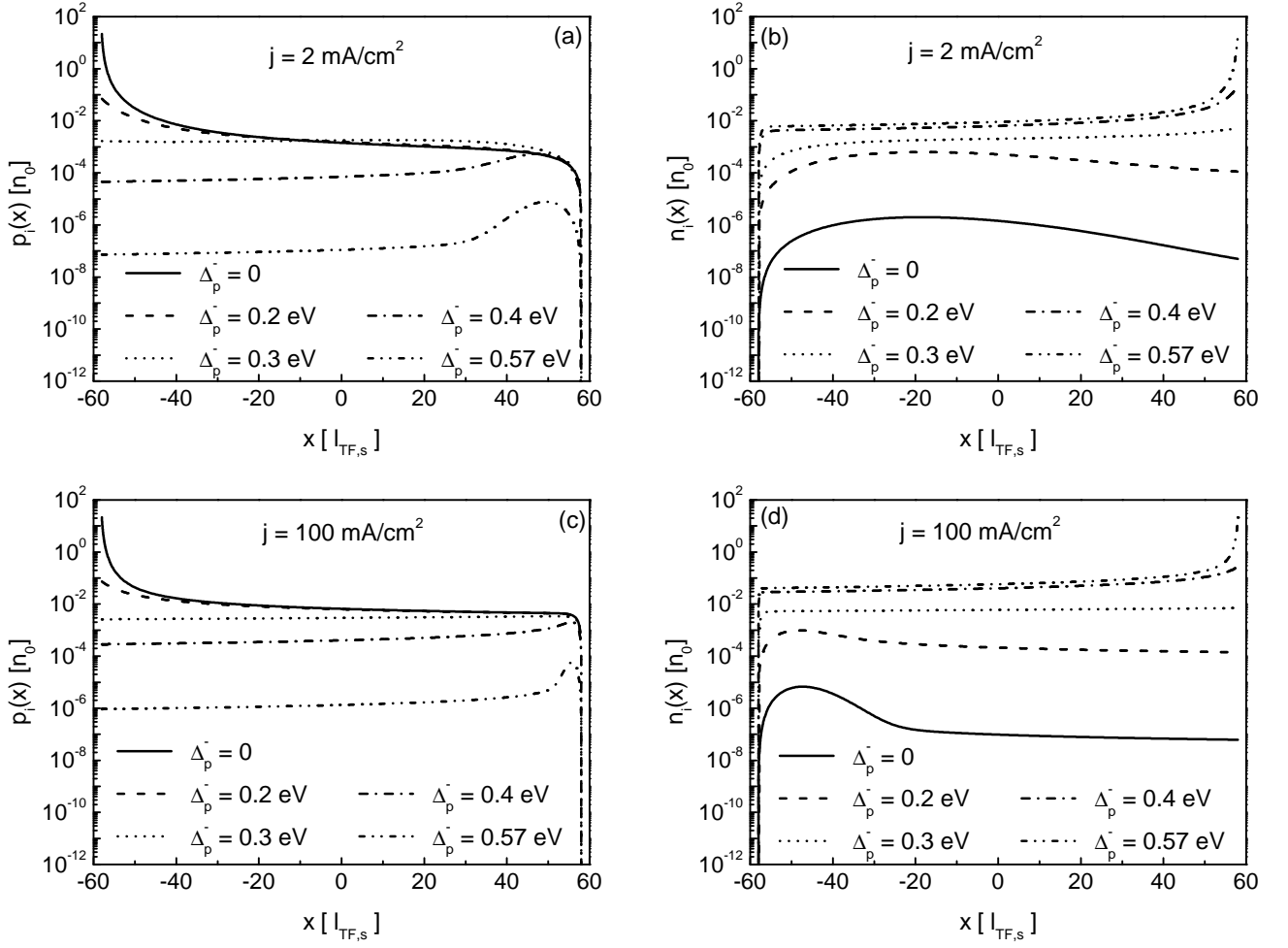}
\caption{Spatial distributions of the charge-carrier densities, $p_i$ and $n_i$, in units of $n_0=\epsilon_i \epsilon_0 kT /
e^2 l_{TF,s}^2$ for a system without recombination for different barrier heights $\Delta_p^-$ and a
constant current density: [(a) and (b)] $j=2$~mA/cm$^2$ and [(c) and (d)] $j=100$~mA/cm$^2$.} \label{fig1}
\end{figure*}

First, we calculate the spatial distributions of charge carriers and
the electric field in the organic layer without accounting for
recombination ($R=0$). The barrier height for hole injection was
varied from 0 to 0.57~eV. For the parameters from
Table~\ref{Materialparameters}, using equations~(\ref{barrier1})
-(\ref{barrier2}) and due to the fixed gap energy of 2.4~eV a hole
injection barrier of 0.57~eV at the anode corresponds to a
barrier-free electron injection from the opposite electrode. In
Fig.~\ref{fig1} the distributions of holes and electrons in the
organic layer are depicted for two different current densities of
$j=2$ and $100$~mA/cm$^2$.
The electron and hole density profiles change slightly with the
current but depend strongly on the respective injection barrier
heights. When one injection barrier is small, a high charge carrier
density appears near the respective electrode decreasing
monotonously through the organic layer and dropping quickly near its
ejecting electrode. In this case the injection barrier at the other
contact is large and thus, the density of the respective charge
carriers is relatively small. Due to the absence of recombination
all injected charge carriers have to traverse the entire organic
layer. The large amount of space charge emerging at the low
injection-barrier electrode prevent the ejection of the traversed
charge carriers injected at the high barrier contact. This gives
rise to a local maxima of their concentration near their ejecting
electrode. The effect is specific as long as no hole-electron
recombination occurs.

\begin{figure*}[!htbp]
\includegraphics{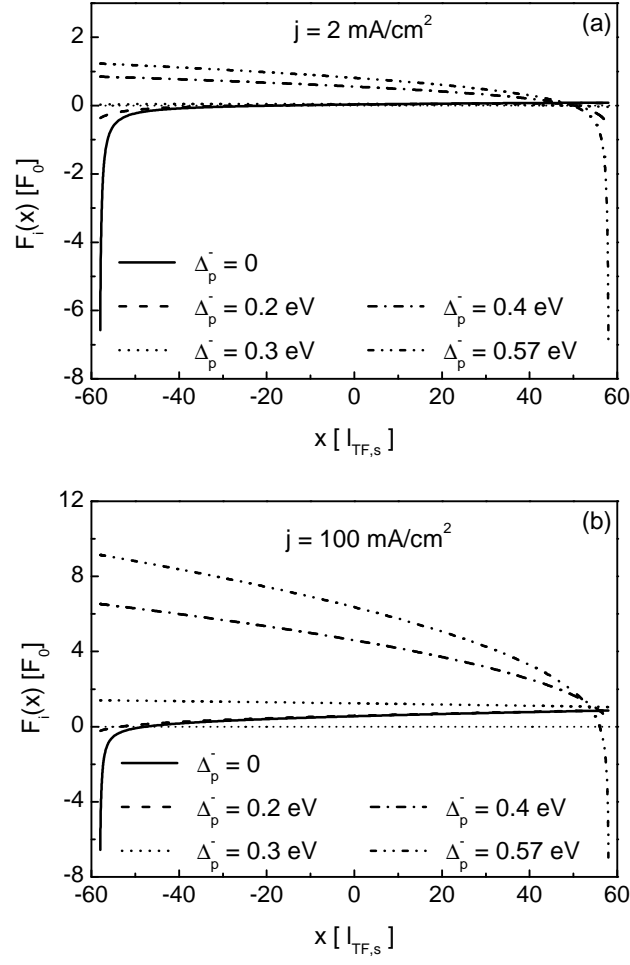}
\caption{Spatial distributions of the electric field $F_i$ in units of $F_0= kT /
e l_{TF,s}$ for a system without recombination for different barrier heights $\Delta_p^-$ and a
constant current density: (a) $j=2$~mA/cm$^2$ and (b) $j=100$~mA/cm$^2$.} \label{fig2}
\end{figure*}

The distribution of the electric field across the organic layer is shown in Fig.~\ref{fig2} for the same
parameter values as in Fig.~\ref{fig1}.
The electric field strongly varies with the change of the injection
barrier. When the heights of the injection barriers at the contacts
differ significantly, one kind of the charge-carrier dominates. As a
result, the electric field changes monotonously between the
electrodes, just like in the single-interface case (see, for
example, Ref.~\onlinecite{PRB2007}).
In the limit of barrier-free
injection the field profile beyond the virtual electrode~\cite{PRB2007} is well described in the frame of the
space-charge approximation.
Once the heights of electron and hole
injection barriers at the cathode and the anode are comparable,
nearly equal hole and electron densities prevail in the organic layer.
Hence, the charge density is small and the electric field appears to
be almost constant with a small maximum in the middle of the organic layer.

\begin{figure*}[!htbp]
\includegraphics{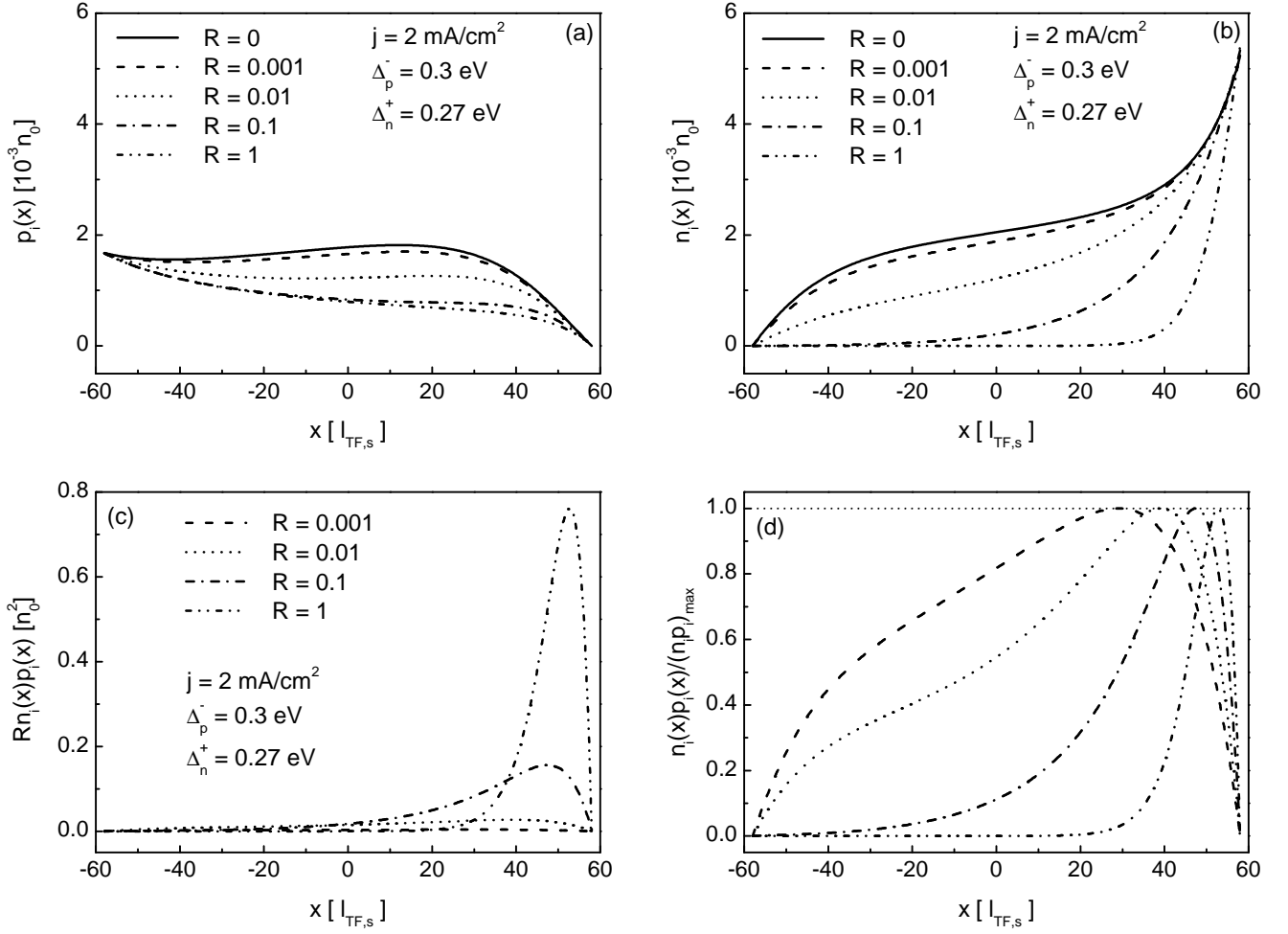}
\caption{Spatial distributions of (a)~the hole density $p_i$, (b)~the electron density $n_i$, (c)~the recombination rate,
and (d)~the recombination zone $\sim$~$n_i p_i$ for a system with the barrier
height $\Delta_p^-=0.3$~eV and the current density $j=2$~mA/cm$^2$ for different values of the recombination
parameter $R$. The carrier densities are measured in units of $n_0=\epsilon_i \epsilon_0 kT /
e^2 l_{TF,s}^2$. The product $n_i p_i$ in the part (d) is normalized on its maximum value.} \label{fig3}
\end{figure*}

\begin{figure*}[!htbp]
\includegraphics{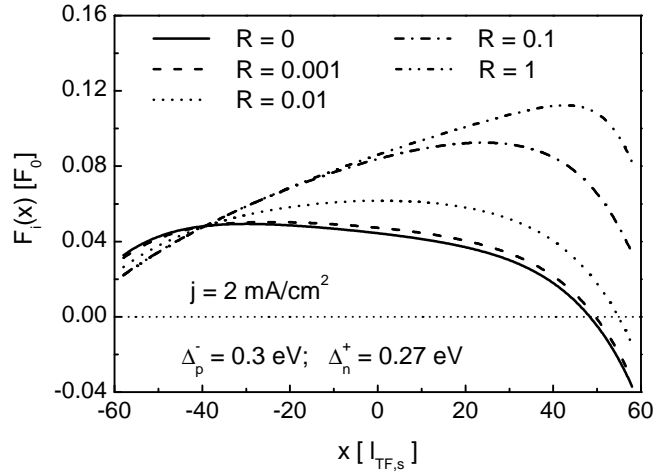}
\caption{Spatial distributions of the electric field
$F_i$ in units of $F_0= kT /e l_{TF,s}$ for a system with the barrier height $\Delta_p^-=0.3$~eV and the
current density $j=2$~mA/cm$^2$ for different values of the recombination parameter $R$.} \label{fig4}
\end{figure*}

Charge recombination gives rise to severe changes in the charge
carrier density and field distributions in the organic layer. First,
let us consider the case of comparable injection barriers, {\it
i.e.} $\Delta_p^-=0.3$~eV and $\Delta_n^+=0.27$~eV. The dependences
of charge-carrier densities on coordinate $x$ are shown in
Figs.~\ref{fig3},a and b for different values of the recombination
parameter $R$.
The annihilation of traversing holes with traversing electrons leads
to a strong spatial decrease in the charge-carrier densities. The
change occurs already for a small increase of the recombination
parameter up to $R=0.1$. A subsequent increase of the recombination
parameter up to 1 influences the picture insignificantly. In
Fig.~\ref{fig3},c the space distribution of the recombination rate $R n_i p_i$ is shown.
To illustrate the position of the recombination zone, which is proportional to the product $n_i p_i$,
the distribution of $n_i p_i$ normalized by its maximum value is shown in Fig.~\ref{fig3},d.
For small $R$ the recombination zone extends almost
over the entire organic layer, but with increase of $R$ it moves
close to the electron-injecting electrode.
This is because of the assumed large difference (two orders of the
magnitude) in the mobilities of the holes and electrons in the
organic material. The injected holes can traverse nearly the
entire organic layer before they recombine with the electrons. The
influence of recombination on the respective electric field
distribution is displayed in Fig.~\ref{fig4}.
Even small values of recombination parameter (up to $R=0.1$) change
the electric field significantly, moving its maximum to the region
of the maximum of the recombination rate. At low $R$ the charge
density in the center of the device is small giving rise to a nearly
constant electric field. At the contacts however, where one charge
carrier specie dominates space-charge effects emerge. Upon increased
$R$ nearly the entire organic layer is dominated by holes and only
close to the cathode a substantial electron density emerges. Hence,
space-charge effects are important in the entire device.

\begin{figure*}[!htbp]
\includegraphics{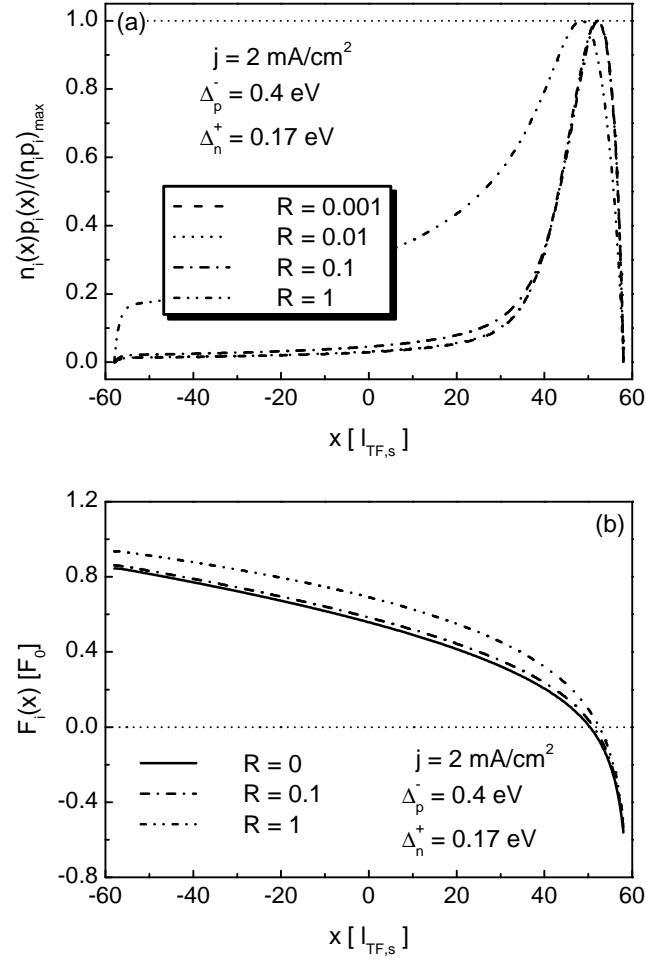}
\caption{Spatial distributions of (a)~the recombination zone $\sim$~$n_i p_i$ and (b)~the electric field
$F_i$ for a system with the barrier height $\Delta_p^-=0.4$~eV and the
current density $j=2$~mA/cm$^2$ for different values of recombination parameter $R$. The electric field is 
measured in units of $F_0= kT /e l_{TF,s}$. The product $n_i p_i$ is normalized on its maximum value.} \label{fig5}
\end{figure*}

\begin{figure*}[!htbp]
\includegraphics{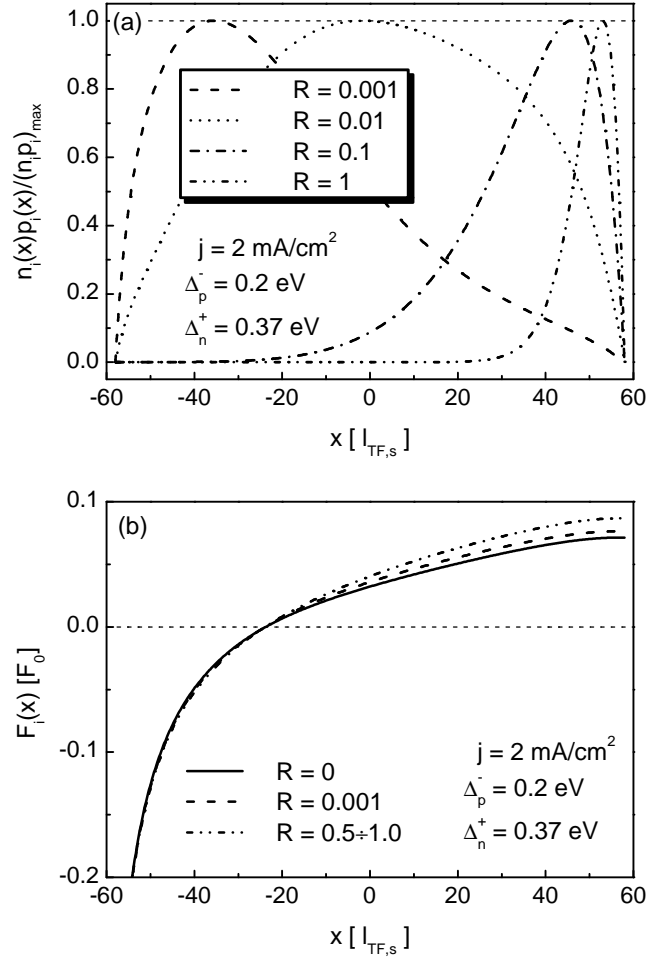}
\caption{The same as in Fig.~\ref{fig5} but for the injection barrier height $\Delta_p^-=0.2$~eV.} \label{fig6}
\end{figure*}

In the case of strongly different injection barriers their influence
becomes more pronounced. This is illustrated in Figs.~\ref{fig5}
and~\ref{fig6}, where the recombination zone and the distribution of
electric field are shown for the barrier heights of
$\Delta_p^-=0.4$~eV and $\Delta_p^-=0.2$~eV, respectively.

For $\Delta_p^-=0.4$~eV the electric field distribution is dominated
by the buildup of negative space charge. Due to the high hole
mobility injected holes traverse the organic layer and recombine
close to the cathode where the maximum electron density prevails.
This holds for all $R$ and thus, the recombination maximum is
situated close to the cathode. For $\Delta_p^-=0.2$~eV, when the
injection of holes is more efficient than the injection of
electrons, the field distribution is dominated by the buildup of
positive space-charge. For very small $R$ the maximum of recombination
zone is situated near the hole-injecting electrode since the
injected electrons can traverse the whole organic semiconductor
without recombining and their density is still locally increased near the hole-injecting electrode.
However, with increased $R$ the recombination
probability increases and the few injected electrons recombine
instantaneously after injection since their recombination rate
exceeds their transport rate. Hence, the maximum of the $n_i p_i$ product
moves to the cathode with increased~$R$.

\subsection{Current-voltage characteristics}

Next, we calculate the current-voltage characteristics of the system under consideration.
Knowledge about the distribution of the electric field gives access to the voltage drop $V$ across
the system for a given current density $j$ and hence, to its {\it IV} characteristics.

The voltage drop at the device is defined as
\begin{align}
V=\int_{-\infty }^{+\infty }F_{d}\left( x\right) dx-V_{bi},   \label{V1}
\end{align}
\noindent where
\begin{align}
F_{d}\left( x\right) =\left\{
\begin{array}{l}
F_{s}\left( x\right) -j/\sigma _{s}, \qquad \qquad \ \ x<-L/2, \\
F_{i}\left( x\right) ,\qquad \qquad -L/2\leq x\leq L/2, \\
F_{c}\left( x\right) -j/\sigma _{c},\qquad \qquad \ \ x>L/2,
\end{array}
\right.   \label{Fd}
\end{align}
\noindent and $-V_{bi}$ is the voltage drop in the case of $j=0$,
{\it i.e.} the built-in potential. Using the system of equations
(\ref{eq1i})-(\ref{eq3i}) and boundary conditions
(\ref{bc1})-(\ref{bc4}), the built-in potential can be obtained
analytically. This calculation is presented in the Appendix.
Similarly to the case of unipolar charge-carrier transport, the
built-in potential is given by the difference of the electrode's
work functions,
\begin{align}
eV_{bi}=E_{A,c}-E_{A,s}.  \label{VBI1}
\end{align}
Integrating the field $F_{d}$, we obtain
\begin{widetext}
\begin{align}
\int_{-\infty }^{+\infty }F_{d}\left( x\right) dx  = l_{TF,s} \left[ \frac{\epsilon _{i}}{\epsilon _{s}}
F_{i}\left( -\frac{L}{2} \right) -\frac{j}{\sigma_s} \right] 
 + l_{TF,c} \left[ \frac{\epsilon _{i}}{\epsilon _{c}} F_i \left(
\frac{L}{2} \right) - \frac{j}{\sigma_c} \right] 
+ \int_{-L/2}^{L/2} F_i\left(x\right) dx.  \label{Fdint}
\end{align}
\end{widetext}
Using the boundary conditions, the voltage may be rewritten in the shorter form
\begin{align}
V=\frac{kT}{e} \ln \left[ \frac{p_{i}\left( -L/2\right) }{%
p_{i}\left( L/2\right)} \right] +\int_{-L/2}^{L/2} F_i\left( x\right) dx.  \label{V2}
\end{align}

\begin{figure*}[!htbp]
\includegraphics{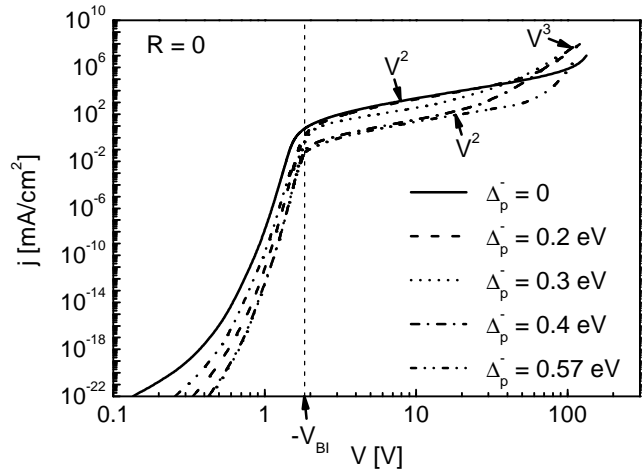}
\caption{{\it IV} characteristics of a system without recombination for
barrier heights $\Delta_p^-=$ 0, 0.2, 0.3, 0.4, and 0.57~eV.} \label{fig7}
\end{figure*}

In Fig.~\ref{fig7} the resulting {\it IV} characteristics are
presented for the case of absence of recombination when the barrier height $\Delta_p^-$
changes from 0 to 0.57~eV.
At voltages $V \lesssim -V_{bi}$ there is no Ohmic-like $j \sim V $
behavior which has been obtained in the cases of the unipolar
transport~\cite{JAP100} or of the bipolar charge injection from
Ohmic contacts~\cite{SyntMet05}. In the calculated range of voltages
(and currents) all {\it IV} characteristics exhibit only a tendency
to this behavior. Hence, it can be supposed that the Ohmic region
lies at even lower voltages. Near the built-in voltage, $ -V_{bi}$,
a relatively wide region starts where, for the cases of barrier-free
injection of carriers (holes or electrons), the SCLC behavior $j
\sim V^2$ of {\it IV} characteristics occurs. This behavior is
similar to that described in the diffusion-free analysis in
Ref.~\onlinecite{Parmenter} and stems from the dominance of
electrons or holes, respectively. When nonzero injection barriers
exist for both kinds of charge carriers, the {\it IV} dependences
differ from the SCLC behavior but all corresponding curves lie between the
curves with $ \Delta_p^-=0$ and $\Delta_p^-=$ 0.57~eV. For high
voltages at all $\Delta_p^-$ the {\it IV} dependences tend to a
transition in the $j ~\sim V^3$ regime which is intrinsic for 
systems with bipolar transport in the recombination-free
case~\cite{SyntMet05}. Yet, in the framework of our model this
region can not be reached at some barrier heights because the used
Boltzmann statistics of carriers in the organic constituent is
violated here.

\begin{figure*}[!htbp]
\includegraphics{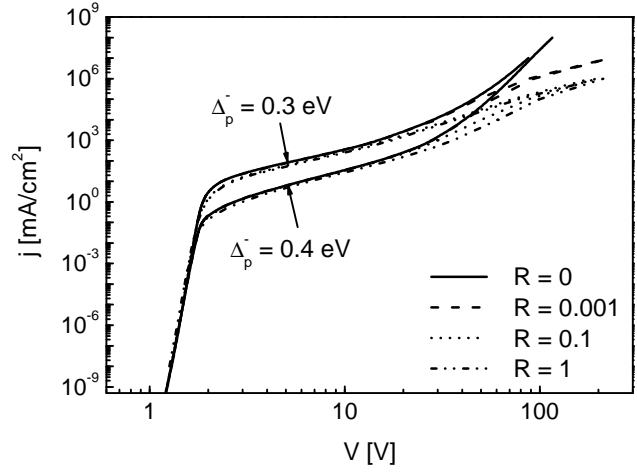}
\caption{{\it IV} characteristics of a system with
barrier heights $\Delta_p^-=$ 0.3 and 0.4~eV for different values of the recombination
parameter $R$.}
\label{fig8}
\end{figure*}

When recombination is accounted for, the {\it IV} characteristics change substantially. This influence is demonstrated
in Fig.~\ref{fig8} on the example of characteristics for the barrier heights of $\Delta_p^-=$ 0.3 and 0.4~eV.
One can see that the most significant effect occurs in the
high-voltage part of the characteristics where the transition from
the $j ~\sim V^3$ regime to the SCLC-like $j ~\sim V^2$ behavior
occurs even for the smallest value of $R=0.001$. At the same time,
the remaining part of the {\it IV} characteristics virtually is not
changed by the recombination. However, it should be noted that with
increase of $R$ the current density slightly increases in the low
voltage part of the {\it IV} characteristics whereas in the region
of the SCLC-like behavior a slight decrease of $j$ is visible.

The significant change of the {\it IV} characteristics at high
voltages can be related to the influence of the recombination on the
modification of the injection barriers. The corresponding
dependences of the effective barriers $\Delta_{\mbox{eff}}^{\pm} $
on the applied voltage $V$ are presented in Fig.~\ref{figEB}.
In general, the effective barriers decrease with the voltage tending
to become equal at vanishing barrier for the recombination-free
case. However, recombination weakens this decrease so that nonzero
effective barriers are obtained at much higher voltages. By that the
difference between $ \Delta_{\mbox{eff}}^{-}$ and $
\Delta_{\mbox{eff}}^{+}$ remains considerable and even increases
significantly for high magnitudes of the recombination rate.
Paradoxically, the prevailing injection barriers provide a {\it
significant} extension of the voltage range where the SCLC-like
behavior of the {\it IV} characteristic occurs.

\subsection{Recombination current and efficiency}

Integrating the product $e B R \left[ n_i \left( x \right) p_i
\left( x \right) - n_{\text{int}}^2 \right]$ [{\it i.e.} the
right-hand side of Eq.~(\ref{difkappa1})] over the thickness of the
organic layer, we obtain a quantity having the dimension of a current density and 
usually called the recombination current density, $ j_r $ (Ref.~\onlinecite{Walker2002}).
It stands for the total number of recombination events in the volume of organic constituent 
per unit square of the interface, per unit time. This quantity 
determines the recombination efficiency of the
device, $ \eta_r = j_r / j$ (Ref.~\onlinecite{Walker2002}). The
calculated dependences of the recombination current density on the
applied voltage are given in Fig.~\ref{fig9} for the barrier heights
$\Delta_p^-=$ 0.3 and 0.4~eV. 
The total recombination
current increases with increased recombination parameter $R$ and
thus a higher luminance of the OLED is expected.
It is clearly seen from Fig.~\ref{fig9} that already for the smallest calculated values
of the recombination parameter $R$ a substantial recombination
current is observed while in the range from $R=0.1$ to 1 the increase in $ j_r $ rather weak.

\begin{figure*}[!htbp]
\includegraphics{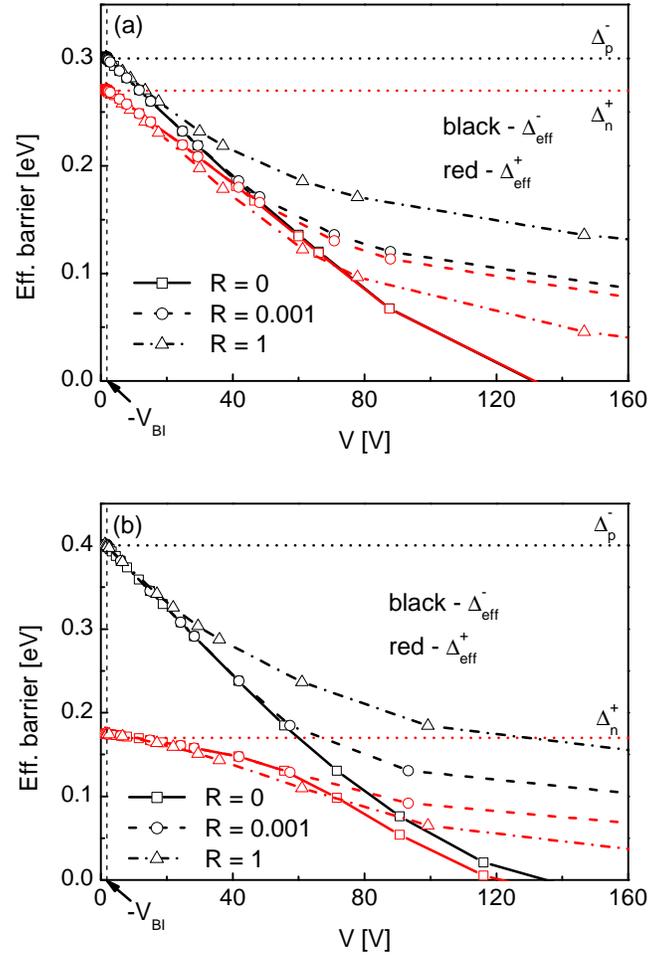}
\caption{Dependence of the effective injection barriers $\Delta_{\mbox{eff}}^{\pm} $ on the applied voltage for the
barrier heights (a)~$\Delta_p^-=$ 0.3~eV and (b)~0.4~eV and for different values of the recombination
parameter $R$.}
\label{figEB}
\end{figure*}

\begin{figure*}[!htbp]
\includegraphics{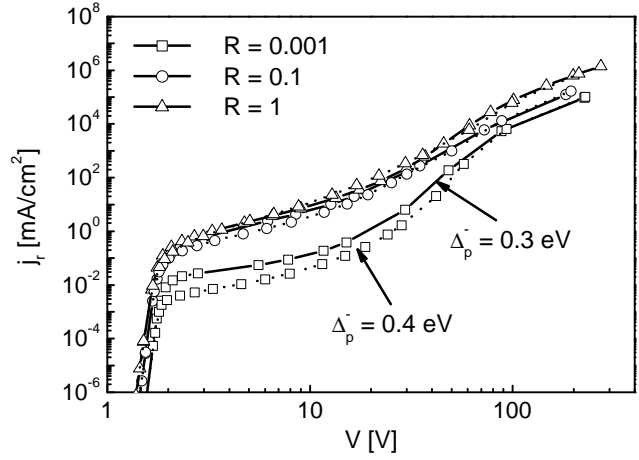}
\caption{Dependence of the recombination current density on the voltage $V$ for a system with
barrier heights $\Delta_p^-=0.3$~eV (solid curves) and 0.4~eV (dotted curves) for different values of the recombination
parameter~$R$.} \label{fig9}
\end{figure*}

\begin{figure*}[!htbp]
\includegraphics{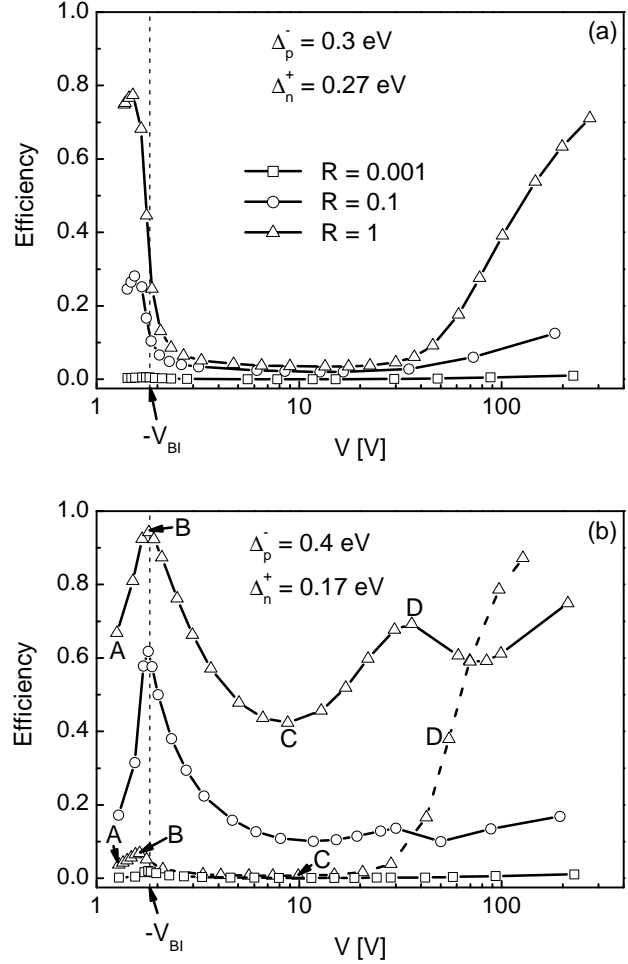}
\caption{Dependence of the recombination efficiency $ \eta_r $ on the voltage $V$ for a system with
barrier heights (a)~$\Delta_p^-=0.3$~eV and (b)~0.4~eV for different values of the recombination
parameter: $R=0.001$ (squares), $R=0.1$ (circles) and $R=1$ (triangles). The $ \eta_r \left( V \right)$
dependence for $\Delta_p^-=0.4 \mbox{~eV}$, equal
carrier mobilities $\mu_{p,i} = \mu_{n,i}$ and $R=1$ is also shown by dashed curve.}
\label{figeff}
\end{figure*}

In general, the recombination currents follow the {\it IV}
characteristics with an increased total recombination rate for an
increased current density. However, the dependence of the
recombination efficiency $ \eta_r $ on the applied voltage unveil
important details. In Fig.~\ref{figeff} the recombination efficiency is depicted
using the same parameters as for the system in Fig.~\ref{fig9}.
The recombination efficiency shows features which have not been discussed in previous works
~\cite{Walker2002,Shen1998,Walker2000,Martin2,Malliaras1,Malliaras2}
where different boundary conditions at the injecting interfaces have
been assumed. In general, a maximum in $ \eta_r \left( V \right)$
is found close to $-V_{bi}$ no matter which injection barriers have
been considered. This may be explained by the detailed analysis of
the carrier densities and partial currents as follows: a fast
increase of the electric current $j$ occurs at voltages $ V \lesssim -V_{bi}$
(Ref.~\onlinecite{JAP100}) corresponding to the Shockley diode
equation for unipolar injection~\cite{Sze}. Assuming a virtually independent
unipolar injection for electrons and holes from the two electrodes one can expect the faster
increase of the recombination current $j_r$ quadratic in carrier
densities which results in the peak close to $-V_{bi}$. At higher
bias $ \eta_r \left( V \right)$ depends strongly on the injection
barrier heights and the charge carrier mobilities. For rather
balanced but high injection barriers ($\Delta_p^-=0.3$~eV) there is
a large range of voltage where the efficiency is strongly attenuated
independently of the recombination parameter. Assuming that the
built up of space-charge is small due to the substantial injection
barriers, the electric field is constant in the device and the
injection barrier lowering scales linearly with the applied voltage
as can be seen from Fig.~\ref{figEB},a in between 3 and 30~V. However, for low
bias the injection barrier lowering is negligible so that the
density of electrons and holes stays roughly constant for a wide
voltage range. Therefore the recombination current does not change
if the bias increases. At the same time, since the total current is
proportional to the electric field a drop of the efficiency upon the
bias increase is expected. For an initiation of barrier lowering,
the charge carrier densities of electrons and holes grow
exponentially with the barrier lowering and thus an increase in the
recombination current sets in. The efficiency saturates and
increases again. For a further increase in bias the reduction of the
injection barrier heights is weakened since space charges emerge.
The respective injection barriers tend to diverge leading to a
higher injection barrier for holes than for electrons balancing the
charge carrier densities in the insulator. Since balanced
charge-carrier densities result in an optimized recombination rate
the self-balancing effect leads to an increase in efficiency.

Introducing asymmetric injection barriers the efficiency-voltage
characteristic changes dramatically. For a higher value of the hole
injection barrier, $\Delta_p^-=0.4$~eV, and thus a low electron
injection barrier the attenuation of the efficiency due to high
injection barriers is missing. The most pronounced change in the
efficiency curve is the appearance of the peak denoted by~D in
Fig.~\ref{figeff},b for $R=1$. Increasing the voltage coming from
$-V_{bi}$ leads to a buildup of space-charge for the charge
carriers with the lower barrier, in this case electrons.  Strong
injection of electrons contributes to the electric current $j$ but
does not add much to the recombination current because the minority
carriers remain in the barrier-limited regime. This provides the
fall of the efficiency down to the point~C. However, since the
electron mobility is low, the negative space charge does not result
in a strong current. On the other hand, the Langevin recombination
parameter $B$ is large since it is dominated by the larger hole
mobility. Therefore, the efficiency is not fully attenuated. In the
area C to~D where the electrons are already in the space-charge
regime with approximately constant and low barrier, the barrier for
holes decreases (see Fig.~\ref{figEB},b) providing exponential
increase of the minority carriers. This does not contribute much to
the electric current $j$ but strongly promotes the recombination
current $j_r$. Further increase of the voltage depresses the
injection barrier at the hole-injecting side leading to an
substantial amount of holes accumulated at the anode. Now the above
mentioned self-balancing effect sets in and the efficiency tends to
increase again. Apparently, the charge carrier mobilities are
important for the efficiency peak~D. Assuming comparable charge
carrier mobilities for electrons and holes, a strong reduction of
the efficiency can be observed below 30~V.

\begin{figure*}[!htbp]
\includegraphics[width=12cm]{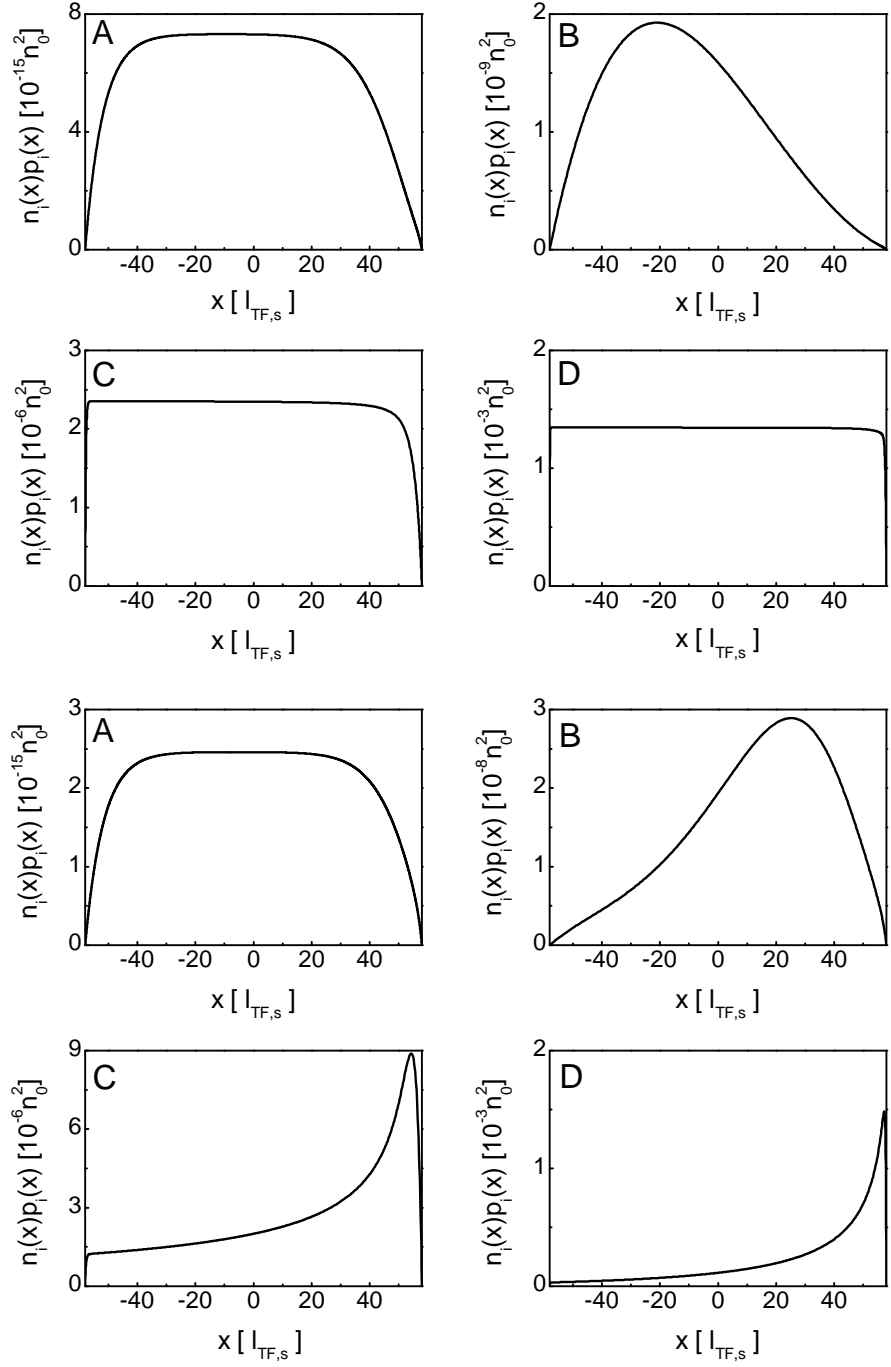}
\caption{Spatial distributions of the recombination zone $\sim$~$n_i p_i$ at different voltage for a system with the
barrier height $\Delta_p^-=0.4$~eV, the recombination parameter $R=1$ and the hole
mobilities $\mu_{p,i} = \mu_{n,i} $ (upper four figures) and $\mu_{p,i} = \mbox{100~} \mu_{n,i} $ (bottom four figures) 
for the points
of the $ \eta_r \left( V \right)$ curves marked in Fig.~\ref{figeff},b by capital letters from A to D. The carrier 
densities are measured in units of $n_0=\epsilon_i \epsilon_0 kT /e^2 l_{TF,s}^2$.} \label{figeff3}
\end{figure*}

It should be also noted that the change of the mobility ratio (for
example, by decrease of the hole mobility) not only varies
quantitatively the values of the current and the efficiency but also
acts strongly on the position and the shape of the recombination
zone. This is seen from Fig.~\ref{figeff3}, where the spatial
distributions of this zone, calculated in several points of the $
\eta_r \left( V \right)$ dependences of Fig.~\ref{figeff},b for the
cases $\mu_{p,i} = \mu_{n,i}$ and $\mu_{p,i} = \mbox{100~} \mu_{n,i}
$, are presented.
With increased applied voltage, the recombination zone extends
through the whole organic layer as long as equal mobilities
$\mu_{p,i} = \mu_{n,i}$ are assumed, whereas in the case of the
large mobility difference there is a pronounced peak of the
recombination zone moving to the electron-injecting electrode since
the injected holes are faster than the electrons.

Thus, additionally to the case of fully balanced injection and
transport properties, which is often difficult to achieve in
single-layer devices~\cite{Walker2002}, it may be possible to
maximize the recombination efficiency in an OLED with imbalanced
injection and transport properties compensating the small amount of
the minority carriers by their relatively high mobility with respect
to the mobility of the dominating carriers.

\subsection{Comparison with experiment}

To evaluate the presented model it is compared to experimental
data obtained from a diode consisting of a single
poly(phenylene vinylene) (PPV) layer of thickness $L=100$~nm
sandwiched between ITO and Ca electrodes. The characteristic
energies of the LUMO and HOMO levels of the employed PPV are 2.8~eV
and 5~eV, respectively and the hole mobility $\mu_{p,i}$ equals $5
\cdot 10^{-7}$~cm$^2$/(V~s) (Refs.~\onlinecite{Blom,Blom2000}). The
ratio of the electron and hole mobilities is assumed to be $
\mu_{n,i}/\mu_{p,i}=0.01$. It is known, that the work function of
ITO is sensitive to the cleaning procedure and thus, it may be varied
from 4.7~eV to 5~eV (up to the energy of the HOMO level). Therefore, 
we are allowed to consider the injection barrier $\Delta_p^-$ as one
of the fitting parameters. The other barrier, $\Delta_n^+$, is
supposed to be unchanged and equals 0.07~eV.

\begin{figure*}[!htbp]
\includegraphics{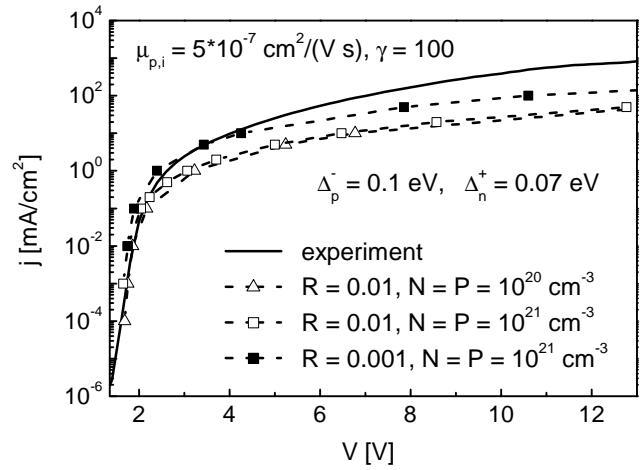}
\caption{Measured {\it IV} characteristic of ITO-PPV-Ca structure (solid curve) and its approximation with
different values of the model parameters (dashed curves with symbols).} \label{fig10}
\end{figure*}

In Fig.~\ref{fig10} the measured {\it IV} characteristic of the ITO/PPV/Ca structure is shown as well as
the {\it IV} dependences calculated for $\Delta_p^-=0.1$~eV (the best fitting value) and for different values
of fitting parameters $R \mbox{ and } N(=P)$.
On the one hand, the calculation satisfactory reproduces the
magnitude and the general form of {\it IV} characteristic using
reasonable parameters. On the other hand, however, the {\it exact} shape 
of the {\it IV} curve can not be fully approximated by a unique set of parameters.
This is due to the fact that the proposed model misses still some
important features of insulators, in particular of organic
semiconductors. First of all, realistic DOS shapes for the LUMO and
HOMO levels as well as trap levels in the
insulator layer have to be accounted for. This leads effectively to
a field and charge carrier density dependence of the mobilities which may
increase the calculated currents at larger voltages. It should be
also noted, that due to this the ratio of the hole and electron
mobilities varies with the change of the applied
voltage~\cite{Blom,Blom2000}.

\begin{figure*}[!htbp]
\includegraphics{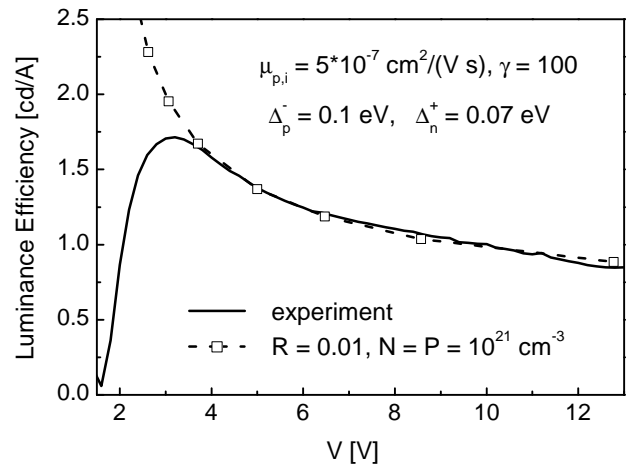}
\caption{Measured voltage dependence of the luminance
efficiency of ITO-PPV-Ca structure (solid curve) and its fitting with the model parameters (dashed curve). The last dependence
is shown in arbitrary units.} \label{fig11}
\end{figure*}

Comparing the luminance efficiency in
Fig.~\ref{fig11} with the calculated recombination efficiency a
good reproduction of the voltage dependence can be obtained above 3~V.
Here, the calculated recombination efficiency was normalized to the
measured luminance efficiency ignoring e.g. the voltage dependence
of the extraction efficiency.  
There is, however, a strong discrepance of the measured and fitted 
efficiency curves at lower voltages.
Again, this difference may be due to the absence of trap levels in our simple model.
At low voltages, strong trapping could arise close to the electrode shifting the 
recombination zone to the injecting contact giving rise to a strong reduction of the 
luminance efficiency due to quenching effects~\cite{Blom2000,BlomAPL97,Blom}.  

Concentrating on the effect of injection on the device performance
one has to consider also a possible formation of the dipole layers
at the electrode/organic interface which may have crucial
consequences for the prevailing charge-carrier injection barriers.
Dipole layers may be a result of self-organization or a consequence
of the deposition of some functional layer at the
interface~\cite{Ishii,Kahn}. The phenomenon of a dipole-layer
formation cannot be incorporated self-consistently in the developed
phenomenological approach, however, an effective change in the
injection barrier heights can be assumed. Notice, that in this case
the magnitudes of injection barriers $\Delta_p^-$ and $\Delta_n^+$
are decoupled and become independent.

If compared to other models of bipolar transport and recombination
in OLEDs the model presented here describes satisfactorily a crossover
from the high-voltage, space-charge limited regime to the
low-voltage, barrier-dominated regime. Even the regime around the
built-in potential which is problematic in theories using
other boundary conditions~\cite{Martin2005,Blom,Arkhipov} is well
described. Comparing with very comprehensive discrete
one-dimensional simulations by Tuti\v{s} {\it et
al.}~\cite{Tutis2001} our approach presents a simpler continuous
description using a strongly reduced number of fitting parameters.

\section{Conclusions}
In the calculation of the charge-carrier transport through
insulators, the fundamental question about boundary conditions
generally arises when charge-carrier injecting interfaces are
involved. In this paper, a simple one-dimensional model describing the bipolar
charge-carrier transport across a semiconductor/insulator/conductor
structure was presented, where the problem of injection was for the first time 
solved self-consistently. 
Essentially, continuity of the electric
displacement and the electrochemical potential was assumed, thus
matching them at the injecting interfaces of the electrodes with the
insulator. 
Additional boundary conditions were defined far
into the electrode materials, where the influence of the involved
materials on each other is negligible so that they can be regarded
as independent. 
Considering the Poisson equation, the electric-field
distributions and the current-voltage characteristics were derived
using as an example the organic semiconductor sandwiched between 
ITO and Ca electrodes. The influence of different injection
conditions and recombination rates on the current transport
properties of such a structure was analyzed. It was found that the
injection barriers determine the dominating carrier type and
strongly influence the distributions of the charge and of the
induced field in the organic layer. The recombination influences on
these quantities most strongly when the heights of injection
barriers for holes and electrons are comparable. It was shown that
the injection barriers and recombination together with the mobility ratio
define the position and shape of the recombination zone in the organic
layer as well as drastically affect the current-voltage
characteristics of the considered diode structure. We have also
found that in the case of significantly different injection barriers
a high recombination efficiency of a diode may be still achieved
in the case of a relatively high mobility of the minority carriers.
Notice, that the controlling  of the carrier mobility in organic devices
is experimentally possible~\cite{Bruetting}.

Finally, we can conclude that the presented self-consistent model of charge injection
reveals unambiguously that injection barriers and recombination rate are the governing factors
controlling the transport properties of semiconductor/insulator/conductor structures. 
On the other hand, this model cannot yet provide a complete description of the particular case 
of OLEDs because some specific features of these systems are still missing.
Most important of them are the realistic DOS shapes including energetically distributed 
trap states~\cite{Lampert,BlomAPL97,BlomIEEE,Blom}
or concentration- and field-dependent carrier mobilities~\cite{Pai} characteristic of organic semiconductors.
The latter feature is followed by the exponential-type dependence of the injection current on the device 
thickness~\cite{Murgatroyd,Malliaras2} which can hardly be expected within our consideration. 
Further extension of the model concerning the inclusion of the above-mentioned features is 
necessary for better application to experimental data and is now in preparation.

\begin{acknowledgments}
This work was supported by the Deutsche Forschungsgemeinschaft through the Sonderforschungsbereich 595.
\end{acknowledgments}

\appendix*
\section{Calculation of the built-in potential in the case of bipolar transport}

The built-in potential defines as
\begin{align}
V_{bi}=\int_{-\infty }^{+\infty } \left. F_{d}\left( x\right) \right|_{j=0} \; dx, \label{VBIA1}
\end{align}
with $F_{d}\left( x\right)$ from Eq.~(\ref{Fd}).
In the case of $j=0$ both current densities $j_p$ (holes) and $j_n$ (electrons) are equal to zero.
From Eqs.~(\ref{currentn2fin})-(\ref{currentp2fin}) we get
\begin{align}
\frac{kT}{e} n_i^{\prime } \left( x \right) + n_i \left( x \right) F_{i}\left( x\right) =0, \\
\frac{kT}{e} p_i^{\prime } \left( x \right) - p_i \left( x \right) F_{i}\left( x\right) =0,
\end{align}
\noindent or, combining these expressions,
\begin{align}
F_{i}\left( x\right) = \frac{kT}{2e} \frac{d}{dx} \ln \left[ \frac{p_i \left( x \right)}
{n_i \left( x \right)} \right], \label{FxJ0}\\
\frac{d}{dx} \ln \left[ p_i \left( x \right) n_i \left( x \right) \right] = 0. \label{pnJ0}
\end{align}
Integrating in Eq.~(\ref{VBIA1}) with $F_{i}\left( x\right)$ from (\ref{FxJ0}), we find
\begin{widetext}
\begin{align}
V_{bi}  = l_{TF,s} F_{s}\left( -L/2 \right) + l_{TF,c} F_{c}\left( L/2 \right) 
 + \frac{kT}{2e} \left\{ \ln \left[ \frac{p_i \left( L/2 \right)}{n_i \left( L/2 \right)} \right] -
\ln \left[ \frac{p_i \left( -L/2 \right)}{n_i \left( -L/2 \right)} \right] \right\}. \label{VBIA2}
\end{align}
\end{widetext}
Using the boundary conditions (\ref{bc1})-(\ref{bc4}) with $j=0$, we directly obtain
\begin{align}
V_{bi} =\frac{1}{e} \left( \Delta_n^+ - \Delta_n^- \right). \label{VBIA3}
\end{align}
\noindent Accounting for Eq.~(\ref{barrier1}), it follows immediately that
\begin{align}
e V_{bi} = E_{A,c} - E_{A,s}. \label{VBIA4}
\end{align}

It should be also noted, that from Eq.~(\ref{pnJ0}) follows that the product $p_i \left( x \right) n_i \left( x \right)$
takes the constant value everywhere in the organic layer.
To provide zero current in the absence of applied voltage and be consistent
with Eqs.~(\ref{eq1i})-(\ref{eq2i}), we choose this constant equal to $n^2_{\text{int}}$.

\end{document}